# A Novel Approach to Compress Centralized Text Data using Indexed Dictionary

Vivek Dimri
Jamia Hamdard, New Delhi
vivdimri@gmail.com

Prof Ranjit Biswas
Jamia Hamdard, New Delhi
ranjitbiswas@yahoo.com

**Abstract-** Data compression is very important feature in terms of saving the memory space. In this proposal, an indexed dictionary based compression is used for text data, where the word's reference in dictionary is used in compression. This approach is not file based; a common dictionary is used for compression. Which contains the words, the position of the word in dictionary is one of the key parts of encoded frame which is compressed form of the text word. This is lossless compression. This compression approach is also take cares of small words like one or two characters words which usually decrease the efficiency of compression algorithms. This approach is also deals with file having special characters as a word. Special character words, alpha numeric words, normal texted words and small words all deals differently which makes this approach more efficient. Since a centralized dictionary is used for data compression, therefore; this approach is not preferred for transfer compressed file, while it is suitable to store text data in compressed form in hard disk drive and centralized storage or cloud drive for memory utilization.

## I. INTRODUCTION

Text compression is typically used to save storage. Dictionary-based text compression is a compression technique, which provides the most significant improvement in the text compression performance [1]. It is based on the notion of replacing whole words with shorter codes from the real text file. Usually large volume of words with their codes is preserved in the dictionary [1]. In indexed dictionary based compression, a common or global dictionary is used for compression. As compare to other compression algorithm or techniques proposed approach is not file based it is indexed dictionary based so once a dictionary has been made it is easier to compress and decompress the text. Using this approach good compression ratio can be achieved. This approach will be very useful if a text file having larger words.

It is important to implement dictionary such a way so addition of new words are easier as well as fetching the words should also not be complex. A smart dictionary is suggested for this approach, like if searching word does not exist in dictionary, that word automatically at the end, against appropriate index. Detail description of compression process using proposed approach is defined in proposed solution section.

LITERATURE REVIEW

A dictionary based method can be improved significantly by suitably reusing parts of the dictionary entries. Extensive simulation results have been used to support the efficiency of the superiority of the proposed technique against the corresponding already known techniques with respect to compression ratio and dictionary size [4].

Recent techniques uses bit toggle information to create matching patterns and thereby improve compression ratio. However, due to lack of efficiency in matching scheme, the existing techniques can match up to three bit differences [1].

The size of the original files which is carefully compared with the compressed files based on the LIPT, StarNT and WRT preprocessing techniques. It is very vivid that bzip2 + StarNT could provide a better compression performance that maintains a convincing compression and decompression speed while it is compared with LIPT [10].

## II. PROBLEM STATEMENT

Since optimum utilization of memory is very important concern and in order to do that, data must be stored in compressed format so it will take less space in memory. So far the techniques which are being used to compressed text data uses character by character compression rather entire word or some other which works on words they only use reference from dictionary.

Dictionary which is being used in dictionary based compression is not self organized this paper suggest a self organized dictionary which also capable of learning the words which previously is not having. Since this approach is used position of word in compression, thus word's position cannot be change once it has been stored in the dictionary. Therefore, words cannot be rearranged in the dictionary.

Another important fact about the dictionary is, dictionary is indexed based so word directly can be fetched from dictionary for example word[Pos].

### III. PROPOSED SOLUTION

The proposed dictionary based compression method will first read the word from text file and then it will check its location (reference) in dictionary and put the specific compressed value as per encoding is define in figure-3.

If the word is not found in the dictionary then it will be inserted to appropriate location as per the construction rule define for dictionary.

**Dictionary:** Proposed dictionary is index based. There are two indexes store the words in dictionary. Since a text file may have normal character data word, alpha numeric and special symbol word, and one or two character long word, in order to deals with all mentioned words following three kinds of dictionary suggested in this approach.

**i. Dictionary for normal text data:** In this dictionary shown in figure-1, two indexes are used to store data. First index or main index is for initial character of word and second index or sub index is total numbers of characters in the word. All the characters of the word must converted either small case or uppercase (depends on developer choice in which case he wants to implement the dictionary) before storing in the dictionary. The range of main index will be 26 because, this dictionary store only character data which are already converted in single case. Second index of the dictionary is called sub index which is the word length (total number of characters in the word), and then at the last word is stored in dictionary in sequential fashion. For example; word "any" has to store in dictionary, for this word, initial character is 'a' therefore, main index of the word would be 1 (1 for a, 2 for b, and so on). Now total number of characters in the word any are 3, therefore; the sub index will be 3, free position available in the following dictionary against main index 1 and sub index 3 is 51, so word any will be stored at position 51 of dictionary with index 1 sub-index 3.

**ii. Dictionary for single and double character words:** Since; single and double characters words are the main cause which decrease the compression ratio of any compression approach and in order to diminish this, a separate dictionary is proposed which will store the single and double character words including special character words as per shown in figure-2. It is the responsibility of compression algorithm to take care of normal character and special character words.

**iii. Special character words:** There is a separate dictionary for storing special character words which are stored in the dictionary as per number of characters in word. There is no main index only single index (sub index) is used to store word. Figure-3 shows the dictionary for special characters word.

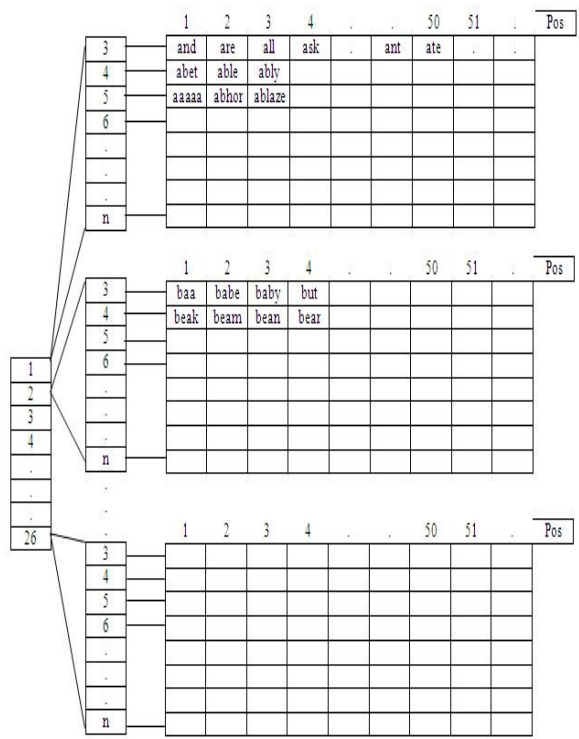

Figure-1: Dictionary for word having more than two characters

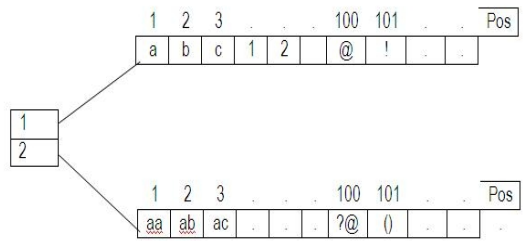

Figure-2 Dictionary for the word having one or two characters

|   | 1 | 2 | 3 | 4 | . | 50 | 51 | Po |
|---|---|---|---|---|---|----|----|-----|
| 3 | a12 | a11 | a@1 | h!S |  |  |  |  |
| 4 | S#2@ | S$S% | b3h6 | 236% |  |  |  |  |
| 5 |  |  |  |  |  |  |  |  |
| 6 |  |  |  |  |  |  |  |  |
| . |  |  |  |  |  |  |  |  |
| . |  |  |  |  |  |  |  |  |
| . |  |  |  |  |  |  |  |  |
| n |  |  |  |  |  |  |  |  |

Figure-3 Dictionary for the word having special characters

Basic operations of dictionary that is insertion and searching word is defined below-

**Insertion:** Insertion operation specifies inserting or adding a new word in dictionary. In order to insert a new word first of all that word has to be converted in either case (upper or lower) and its associated bits must be assigned in encoding tag (defined in encoding section later). Once case of the word (i.e. upper or lower) has changed as per dictionary rule; the first character of the word is find out and also count the number of characters in the word. Initial character will be the first index and number of characters in the word will be the second index. Word will be stored in the dictionary at the end as per number of characters in the words.

For example: suppose word Bulk is not present in the dictionary then first of all, it will be converted to either uppercase or lower case (depends in which case directory is been implemented) then we will find its index value which is 2 (2 for b) and the number characters in the word Bulk i.e. 4, then it will be stored against sub-index 4 of main index b. The word will be stored at the end of the row.

Searching word: Word can be search as per index value given for the compressed string. For example suppose from above mention dictionary, word against indexes 2,4,2 have to find, for this, main index is 2 and sub index is 4 and the location of word in dictionary is 2, that is the word beam.

**Compression:** Encoding format for the compression of word using index based dictionary is given below:

| F | NC | CS | Ic | Pos |            |
|---|----|----|----|----|------------|
| 2 | 4  | 0-16 | 5 | 8 | No of bits |

Figure- 4 Encoding format for compression

**Description of fields:**

**Flag or F:** This 2 bits field is used to specify whether accepted word is space, newline, special characters or simple word. For a space and newline last three fields (CS, Ic, Pos) are discarded.

**Number of Characters or NC:** This 4 bit field specifies how many characters are there in a word. This field is used as sub index for word in the indexed dictionary.

**Case Sensitivity or CS:** This field describes the case bit of each character of word. For lower case dictionary implementation, if character of the word is in upper case then it will have value 1 at that position else 0. For example; word "ThiS" will have the CS value 1001 since it's first and last characters are in upper case and middle characters are in lower case. This field is only applicable if flag is 11 otherwise this field is discarded.

**Initial Character or Ic:** This field contains initial character of the word. It is 1 byte (8 bits) long field and is worked as main index for the word in the dictionary.

**Position or Pos:** This 8 bits long field specifies the location of word in dictionary.

| Field | No of Bits | Bit Sequence | Purpose |
|-------|------------|--------------|---------|
| F | 2 | 00 | Space |
|   |   | 01 | New Line |
|   |   | 10 | Word with special characters |
|   |   | 11 | Alpha characters word only |
| NC | 4 | - | Specify the number of characters in the word |
| CS | 0-8 | - | Case sensitivity bits |
| Ic | 5 | - | Initial character |
| Pos | 8 | - | Position in dictionary |

Table-1 field description of encoding format of compression

**Compression Techniques:** To compress text, Text file or text data will be compressed as per following steps-

*Step-1* Read the word from file and check it's characters
*Step-2* If all characters are alpha then set flag=11, else set flag as per table-1
*Step-3* If flag=11; Change the case of string to upper or lower case (based on case of dictionary)
*Step-4* Then, find out the initial character and number of characters
*Step-5* Set flag as per table-1
*Step-6* check flag is-

00 or 01 then, set NC= number of consecutive spaces/ newlines and discard CS, Ic and Pos field

10 then set NC= number of special characters and discard CS field

11 then, check
    If NC is less than 3, then,
       Set field CS, Pos and discard Ic field.
    If NC is equal to greater than 3, than
       set NC= number of characters in word and also set each bit of CS field as per case sensitivity of character (1 for upper and 0 for lower) also set Ic flag with initial character and Pos to position of word in dictionary

**Step-7** Write compressed word in file.
**Step-8** Read next word till End of File and goto step-1

**Note:** If special character (like comma, hyphen, etc) is encountered followed by white space, then it is considered as an end of the word. This word special character followed by white space is considered as a two character word and will be stored in dictionary for special character word.

**Decompression of compressed word:**

The data which is compressed by above mention scheme decompressed by following steps-
**Step-1** Read the compressed data from compressed file
**Step-2** Check the flag bit combination, if F is-

00 or 01 then, Put NC number of consecutive spaces/ newlines as per table-1

10 then, put the word which is at the location of Pos in dictionary against main index (Ic field) and sub index (NC).

11 then, Check
If NC is less than 3, than,
    Find the word of index NC at position Pos of dictionary and convert the characters of word to upper or lower case as per CS bits

If NC is equal to greater than 3, than
    Find the word at position Pos of sub index (NC) of index (Ic) and convert the characters of word to upper or lower case as per CS bits

**Step-3** Write appropriate word: word[Ic][NC] into the file
**Step-4** Read next compressed word till End of File and goto step-1

## IV. RESULT

Compression result of the text file poem.txt using proposed method and also comparison with other existing approaches are given below.

*poem.txt*
Friends are far, friends are near,
Friends will be there to lend an ear,
They listen, laugh, and care,
But most of all, they're always there,
Through thick and thin, up and down,
Your true friends are always around,
For treats, hugs and real big smiles,
They'll travel to you from several miles,
They'll always be there to hold you tight,
Anytime, no matter if it's day or night,
You really know when your friends are sincere,
When they always show up to lend their ear.
by- Bea Williams

| S. No | Approach | Original Size | Compressed Size | Compressed % | Compression Ratio |
|---|---|---|---|---|---|
| 1 | WinRAR | 574 | 381 | 33.62 | 0.6637 |
| 2 | 7Zip | 574 | 425 | 25.96 | 0.7404 |
| 3 | GZip | 574 | 274 | 52.26 | 0.4773 |
| 4 | LZW | 574 | 370 | 35.54 | 0.6446 |
| 5 | Proposed | 574 | 298 | 48.08 | 0.5191 |

Table-2 Comparison Table

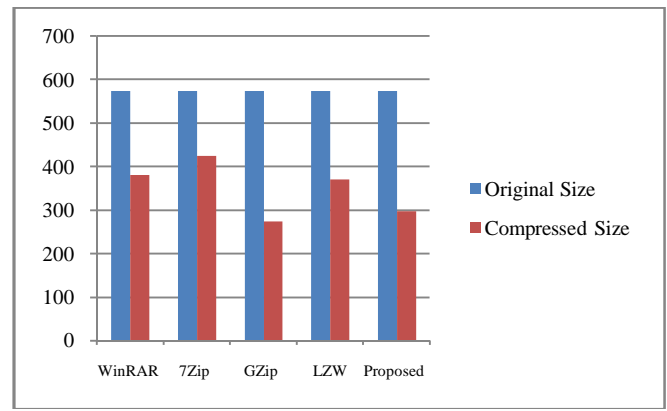

Figure-4 Comparison in terms of compressed size

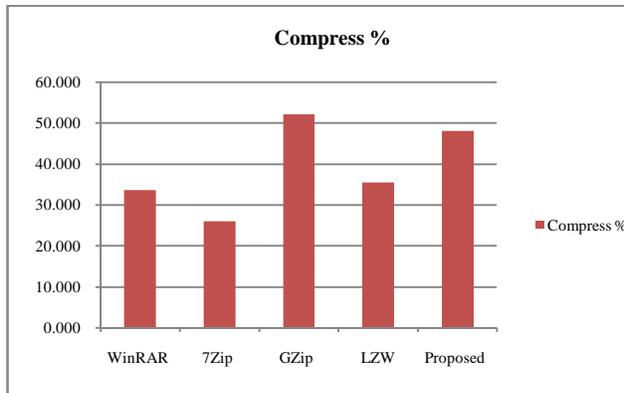

Figure-5 Comparison in terms of compression percent

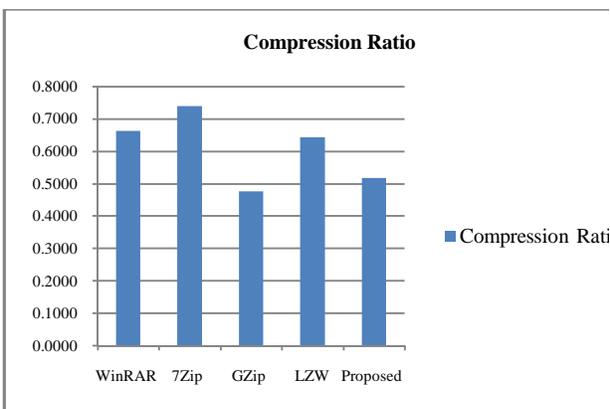

Figure-6 Comparison in terms of compression ratio

## V. CONCLUSION & FUTURE WORK

It is clearly shows by result that, proposed approach is better as compare to other non-dictionary based compression approach. In future, centralized dictionary can be implemented as self organized dictionary using artificial intelligence. This approach can implement in cloud for data storage in compressed form.